\documentclass[11pt,a4paper]{article}
\usepackage[T1]{fontenc}
\usepackage{lmodern,microtype,amsmath,amssymb,amsthm,mathtools}
\usepackage[margin=1in]{geometry}
\usepackage{booktabs,tabularx,array,float}
\usepackage{xcolor}
\usepackage[colorlinks=true,linkcolor=blue!50!black,citecolor=blue!50!black,urlcolor=blue!50!black]{hyperref}
\hypersetup{pdftitle={Self-Reference and Relative Incompleteness: From Logical Obstructions to Hardness Hypotheses},pdfauthor={Jie Luo and Ke Xu}}
\newtheorem{theorem}{Theorem}[section]
\newtheorem{proposition}[theorem]{Proposition}
\newtheorem{lemma}[theorem]{Lemma}
\newtheorem{corollary}[theorem]{Corollary}
\theoremstyle{definition}
\newtheorem{definition}[theorem]{Definition}
\newtheorem{example}[theorem]{Example}
\theoremstyle{remark}

\newcommand{\PA}{\mathsf{PA}}
\newcommand{\N}{\mathbb N}
\newcommand{\Sol}{\operatorname{Sol}}
\newcommand{\Sat}{\operatorname{Sat}}
\newcommand{\Unique}{\operatorname{Unique}}
\newcommand{\Red}{\operatorname{Red}}
\newcommand{\Rep}{\operatorname{Rep}}
\newcommand{\Payload}{\operatorname{Payload}}
\newcommand{\Prov}{\operatorname{Prov}}
\newcommand{\Diag}{\operatorname{Diag}}
\newcommand{\supp}{\operatorname{supp}}
\newcommand{\code}[1]{\ulcorner #1\urcorner}

\newcommand{\Th}{\operatorname{Th}}
\newcommand{\one}{\mathbf 1}
\title{Self-Reference and Relative Incompleteness:\\From Logical Obstructions to Hardness Hypotheses}
\author{Jie Luo \qquad Ke Xu\\[0.5em]
\small State Key Laboratory of Complex and Critical Software Environment,\\
\small Beihang University, Beijing 100083, China\\
\small \texttt{\{luojie, kexu\}@buaa.edu.cn}}
\date{\today}

\begin{document}
\maketitle
\begin{abstract}
We define \emph{relative incompleteness} and use it to provide a unified framework for G\"odelian incompleteness and the limits of local observations in finite SAT. A background is incomplete relative to a target question if two realizations compatible with that background give opposite answers. G\"odelian diagonalization yields a sentence whose truth value differs across two models of the same theory, while solution-dependent blocking yields two formulas that agree on the observed clauses but differ in satisfiability. Both constructions realize a common diagonal relation between truth and acceptance that forces a sound partial evaluator to leave the target undecided.

For a satisfiable formula, replacing a single unobserved clause without changing its variable support can make the formula unsatisfiable exactly when all solutions remaining after deletion of that clause agree on its support. If the original formula has a unique solution, this condition is equivalent to redundancy of the clause. More generally, an observation of a satisfiable formula determines satisfiability exactly when it includes a position at which the formula differs from each unsatisfiable completion. For uniquely satisfiable formulas, the positions that every sufficient observation must include are precisely the redundant ones. We also arithmetize solution blocking and construct a tagged fixed point that refers to its own blocking data. Finally, we use self-referential $K$-SAT to illustrate how relative incompleteness can provide a logical foundation for formulating a quantitative hardness hypothesis.
\end{abstract}

\section{Introduction}
When do the facts already available suffice to answer a question? If two complete situations agree with all those facts but give opposite answers, the answer cannot follow from the facts alone. The issue studied here is how a construction guarantees such a failure for every background in a specified class. Taking all sentences true in the standard natural numbers as axioms gives the complete theory $\Th(\N)$. It is not computably enumerable. G\"odel's theorem instead applies to effectively axiomatized theories of sufficient arithmetic strength, under suitable consistency or soundness assumptions. In the finite setting, a complete input likewise determines satisfiability, but the blocking construction below leaves every permitted local observation compatible with opposite answers. The restrictions differ: effective axiomatization in one case, access to part of an input in the other. The common requirement is a construction that guarantees opposite realizations throughout a specified class of backgrounds.

G\"odel's construction makes a sentence assert its own unprovability in a specified theory~\cite{Godel}; unique-solution blocking makes a formula's unique satisfying assignment determine a constraint that excludes that assignment~\cite{Fang}. Under their respective hypotheses, both constructions leave the retained background unable to decide a target assertion: the theory cannot decide the G\"odel sentence, and the local observations cannot decide the satisfiability of the complete formula. For each eligible theory, the sentence is constructed from that theory's proof predicate. For each permitted observation of a fixed mother formula, an unobserved clause can be replaced while reversing satisfiability. We ask how these constructions realize a common relation between semantic truth and acceptance, and when the finite realization can be carried out by a local edit that preserves the observation.

We adopt a metamathematical viewpoint by taking theories, observation backgrounds, and their capacity to support correct judgments as objects of study. The question is which conclusions follow from a specified background, even when each complete realization has a definite target value. By identifying what a prescribed body of information leaves undecided, relative incompleteness provides a logical starting point for formulating hardness hypotheses.

\subsection{A finite example}
\begin{example}\label{ex:small}
Suppose that a formula has four indexed clauses on variables $x,y$, each using both variables. The first three clauses have been observed:
\[
 Q=(x\lor y)\land(x\lor\neg y)\land(\neg x\lor y).
\]
They exclude $(0,0)$, $(0,1)$, and $(1,0)$, respectively, leaving the unique assignment $(1,1)$. The fourth clause has not been observed. Two possible complete formulas are
\[
 F=Q\land(x\lor y),\qquad
 F^-=Q\land(\neg x\lor\neg y).
\]
The first retains the solution $(1,1)$; the second excludes it and is unsatisfiable. Both have the same first three clauses and the same variable support at every position.
\end{example}
The observation has determined the only possible solution, but it has not determined whether the unobserved clause permits that solution. No consequence of the observed facts can distinguish $F$ from $F^-$, because both complete formulas satisfy those facts. Reading the fourth clause resolves the question.

The replacement is selected from the mother formula's own solution: $(1,1)$ determines the clause $\neg x\lor\neg y$. The original fourth clause repeats the first, so deleting it creates no new solutions. These two facts explain why the replacement reverses satisfiability. Its unobserved position explains why the available information is preserved.

We call this failure of determination \emph{relative incompleteness}. The same definition applies to an arithmetic theory whose models disagree on a sentence and to a local SAT observation whose compatible completions disagree on satisfiability. Some partial observations do decide satisfiability: seeing both unit clauses $x$ and $\neg x$ is enough to establish unsatisfiability, regardless of the remaining clauses. Our task is to identify constructions that guarantee opposite completions for every observation in an explicitly specified class. A single ambiguous observation illustrates the mechanism; the general result must establish the universal quantifier.

\subsection{Background and related work}
The arithmetic account begins with G\"odel~\cite{Godel}. Li~\cite[Sections 5.1 and 5.3]{Li} develops the passage from a proposition-level expression $X\leftrightarrow B(X)$ to a formula concerning the numerical code of a sentence, and explains why the fixed-point construction is a finite substitution procedure rather than a circular definition. Boolos, Burgess, and Jeffrey~\cite{BBJ} provide the logical background on representability, completeness, and incompleteness used below.

Lawvere~\cite{Lawvere} unified diagonal arguments through evaluation and fixed points. Yanofsky~\cite{Yanofsky} gives a set-theoretic presentation and numerous applications. In its elementary form, the construction starts with $e:A\times A\to Y$ and a fixed-point-free map $\alpha:Y\to Y$, and forms $a\mapsto\alpha(e(a,a))$. The resulting function is not represented by a column of $e$. The formulation applies to finite as well as infinite sets. We give an explicit evaluation map placing the two constructions in this set-theoretic pattern. The finite problem then concerns its implementation: which solution structures permit a semantic reversal by changing a single clause position while preserving a specified observation?

A second line of work asks what an arithmetic sentence must do to ascribe a property to itself. Halbach and Visser~\cite{HV} examine this question for arithmetic self-reference. Grabmayr, Halbach, and Ye~\cite{GHY} study constraints on diagonal operators and coding that distinguish paradigmatic self-reference from accidental diagonal sentences. These analyses are relevant to the tagged diagonal construction in Section~\ref{sec:comparison}, where reference to a sentence's code is used to retrieve fixed blocking data.

Popescu and Traytel~\cite{PT} develop abstract conditions for G\"odel's incompleteness theorems and verify them in Isabelle/HOL. Their separation of syntax, representability, and semantic assumptions illustrates why a common diagonal step must be accompanied by conditions establishing its realization. Here the finite realization is governed by an edit restriction and a projection of the solution set.

The solution-blocking construction comes from self-referential $K$-SAT~\cite{Fang}. We use its deterministic ingredients and prove the implications needed here. The probabilistic analysis that supplies suitable logarithmic-width mother formulas remains in that work. Our argument applies to every mother formula satisfying the stated uniqueness, redundancy, and observation hypotheses.

\subsection{Results and organization}
The shared proof step is Lemma~\ref{lem:anti}: an object whose truth value is the complement of its acceptance indicator cannot be decided by a sound partial evaluator. Section~\ref{sec:godel} uses G\"odel's fixed point to show that each theory $T$ in the stated class is incomplete relative to its G\"odel sentence $G_T$. Section~\ref{sec:blocking} uses solution blocking to show that the local observation theory $\Gamma_S(F)$ is incomplete relative to the satisfiability assertion $A_H$ whenever $S$ omits a position at which blocking is possible.

The finite results identify exactly when this implementation is possible. Theorem~\ref{thm:projection} characterizes single-clause blocking by a singleton projection of the retained solution set. Theorem~\ref{thm:implementation} makes this the necessary and sufficient condition for diagonal reversal by one unobserved, support-preserving edit. For unique solutions, the condition reduces to redundancy. Theorem~\ref{thm:observations} characterizes all sufficient observations by their intersections with the difference sets of unsatisfiable completions. Corollary~\ref{cor:indispensable} identifies the positions that every sufficient observation must include.

The quantifiers are central. Every sound computably enumerable extension $T$ of $\PA$ has an independent sentence $G_T$. For each fixed uniquely satisfiable $F$, every observation leaving a redundant position unseen has an unsatisfiable completion, so the same satisfiability assertion remains independent of every such observation theory. The arithmetic target varies with the theory; the finite target stays fixed while the companion may vary.

Section~\ref{sec:discussion} uses Li's distinction between a formal language and its meta-language environment to clarify the self-recognition obstacle. This distinction also frames algorithm design as a mathematical activity and the study of problem complexity as metamathematical. Relative incompleteness gives this distinction a concrete form by identifying a logical obstruction shared by every method restricted to the same information. We then use self-referential $K$-SAT to illustrate how a provable information obstruction can motivate a quantitative hardness hypothesis.

Section~\ref{sec:framework} supplies the semantic framework and common proof step. Section~\ref{sec:comparison} relates both constructions to the set-theoretic diagonal pattern, expresses solution blocking in arithmetic, and constructs a self-referential sentence that reads its blocking data from its own code. The elementary diagonal obstruction is inherited from the established framework; the finite results specify its realization under observation and edit constraints. The tagged arithmetic sentence serves a separate purpose: it refers to its own stored blocking data.

\section{Contexts and Relative Incompleteness}\label{sec:framework}
\subsection{Contexts and realizations}
The example separates three objects: the facts retained, the complete formulas compatible with them, and the question of satisfiability. We call these a \emph{context}, its \emph{realizations}, and a \emph{target}. For an arithmetic theory, the context will be the theory, its realizations will be models, and the target will be the truth value of a sentence. The following definition records only these shared features.

\begin{definition}\label{def:frame}
A \emph{semantic context system} is a tuple
\[
 (\mathcal X,\mathcal I,\mathcal C,P),
\]
where $\mathcal X$ is a set of realizations, $\mathcal I$ is a set of contexts, $\mathcal C(I)\subseteq\mathcal X$ is the realization set of $I$, and $P:\mathcal X\to\{0,1\}$ is a target predicate. A context is \emph{admissible} if $\mathcal C(I)\ne\varnothing$. It is \emph{separating for $P$} if
\begin{equation}\label{eq:split}
 \mathcal C(I)\cap P^{-1}(0)\ne\varnothing,
 \qquad
 \mathcal C(I)\cap P^{-1}(1)\ne\varnothing.
\end{equation}
A pair witnessing~\eqref{eq:split} is called a \emph{semantic separation over $I$}.
\end{definition}
In Example~\ref{ex:small}, $I$ records the first three clauses and all four supports, $\mathcal C(I)$ consists of the possible fourth-clause sign patterns, and $P$ asks whether the resulting formula is satisfiable. The formulas $F$ and $F^-$ witness~\eqref{eq:split}.

There are two useful presentations. For an observation map $V:\mathcal X\to\mathcal Y$, take contexts $y\in V(\mathcal X)$ and $\mathcal C(y)=V^{-1}(y)$. The set $V^{-1}(y)=\{x\in\mathcal X:V(x)=y\}$ consists of all realizations compatible with the observation $y$; in standard terminology, it is the \emph{fiber} of $V$ over $y$. In the finite formula setting, its members are the complete formulas compatible with the observed clauses, which we call compatible completions. For a logical language, take sets of sentences as contexts and models as realizations. In the arithmetic application, countable models with domain $\N$ suffice: completeness and downward L\"owenheim--Skolem provide such presentations of models of consistent extensions of $\PA$. Their arithmetic operations need not be the standard operations on $\N$.

\begin{proposition}[Determination by a context]\label{prop:determination}
For an admissible context $I$, its information determines $P$ exactly when $P$ is constant on $\mathcal C(I)$. In the observation presentation, $P=h\circ V$ for some function $h:V(\mathcal X)\to\{0,1\}$ exactly when, for each attainable observation, all compatible realizations have the same target value.
\end{proposition}
\begin{proof}
The first assertion follows because the nonempty set $P(\mathcal C(I))$ is a singleton exactly when it does not contain both $0$ and $1$. For the second, a factorization makes $P$ constant on each set $V^{-1}(y)$. Conversely, when $P$ is constant on each such set, define $h(y)$ to be that common value on $V^{-1}(y)$.
\end{proof}

Proposition~\ref{prop:determination} concerns an entire class of decision rules. If two realizations compatible with the same observation have opposite target values, no rule depending only on $V$ can be correct throughout $\mathcal X$. Conversely, the existence of a factorization $P=h\circ V$ expresses semantic sufficiency; the computability and resource requirements of $h$ are further properties to analyze. For a finite observation space, $h$ can be represented by a finite table, with construction, storage, and uniform computation across instance sizes posing further resource questions. A complete formula determines its satisfiability; determining that value efficiently is a further question.

Deducing further facts from the same background cannot remove either member of such a pair: every valid consequence must hold in both. To express this formally, suppose contexts are sets of conditions interpreted on $\mathcal X$. Write
\[
 \operatorname{Con}(I)=\{\varphi:\text{every }x\in\mathcal C(I)
 \text{ satisfies }\varphi\}.
\]
\begin{proposition}[Preservation under consequence]\label{prop:closure}
If $I\subseteq J\subseteq\operatorname{Con}(I)$, then $\mathcal C(J)=\mathcal C(I)$. In particular, every semantic separation over $I$ persists after adjoining any collection of its valid consequences.
\end{proposition}
\begin{proof}
The inclusion $I\subseteq J$ gives $\mathcal C(J)\subseteq\mathcal C(I)$. Every realization of $I$ satisfies every member of $J$, giving the reverse inclusion.
\end{proof}
Thus a context and its valid consequences have the same realizations. Sound reasoning cannot exclude any realization of the original background, although it can make an implicit consequence explicitly available. This is the semantic sense in which deduction adds no information here. Deriving an implicit consequence is a computational task whose cost can be studied separately. We use the preservation property for theories and observation records alike.

\subsection{Relative incompleteness}
\begin{definition}
An admissible context $I$ is \emph{incomplete relative to $P$} if it admits a semantic separation for $P$. For a consistent classical theory $\Gamma$ and a sentence $A$, we call $\Gamma$ \emph{incomplete relative to $A$} if
\[
 \Gamma\nvdash A\qquad\text{and}\qquad\Gamma\nvdash\neg A.
\]
\end{definition}
\begin{proposition}[Semantic and deductive formulations]\label{prop:relative}
In classical propositional or first-order logic, a consistent theory $\Gamma$ is incomplete relative to $A$ if and only if it has models satisfying $A$ and models satisfying $\neg A$.
\end{proposition}
\begin{proof}
Soundness makes the two models incompatible with a proof of either alternative. Conversely, if neither alternative is provable, the deduction theorem makes both $\Gamma\cup\{A\}$ and $\Gamma\cup\{\neg A\}$ consistent. Completeness supplies their models.
\end{proof}
This terminology concerns the information in a theory relative to a target. Completeness of the underlying deductive calculus supplies the equivalence with opposite models. A proper part of a description may already determine a particular target; relative incompleteness is characterized by the existence of opposite realizations.

\subsection{Guarantees across a class of backgrounds}
A separation over one context and a guarantee covering a whole class are different assertions. To compare the latter, let $\mathcal B$ index permitted contexts $I_b$ and targets $P_b$ in a semantic context system, allowing the target to depend on $b$. The required statement is
\begin{equation}\label{eq:class-guarantee}
 \forall b\in\mathcal B\ \exists X_b^+,X_b^-\in\mathcal C(I_b):
 \quad P_b(X_b^+)=1,\qquad P_b(X_b^-)=0.
\end{equation}
In the arithmetic application, $b=T$ and the construction supplies $P_b=P_{G_T}$. In the finite application, $b=S$, the context records the observed signs, and $P_b$ is always satisfiability. The formula $F$ supplies the same positive realization for every $S$; only the negative companion need vary.

Equation~\eqref{eq:class-guarantee} records the proof obligation for a guarantee across a class of backgrounds. The reference mechanism specifies how that guarantee is realized. What requires analysis in the present applications is how reference to a theory's proof predicate or a formula's own solution supplies the target or opposing realization, and why the construction works for every eligible background. The next two sections establish these guarantees separately.

\subsection{Truth and acceptance}
A partial evaluator may accept, reject, or leave a question undecided. Its domain is the available record, not necessarily the object itself. Let $V:\mathcal X\to\mathcal I$ and $P:\mathcal X\to\{0,1\}$. A map $D:\mathcal I\to\{1,0,\bot\}$ is \emph{sound} if, for every $x$,
\[
 D(V(x))=1\ \Longrightarrow\ P(x)=1,
 \qquad D(V(x))=0\ \Longrightarrow\ P(x)=0.
\]
Write $a_D(I)=\one[D(I)=1]$. Thus rejection and nondecision both have acceptance indicator zero. No effectiveness assumption is imposed on $D$.

\begin{lemma}[Truth opposite to acceptance]\label{lem:anti}
If $D$ is sound and an object $x_D$ satisfies
\begin{equation}\label{eq:anti}
 P(x_D)=1-a_D(V(x_D)),
\end{equation}
then $D(V(x_D))=\bot$ and $P(x_D)=1$.
\end{lemma}
\begin{proof}
Acceptance would imply $P(x_D)=1$ by soundness and $P(x_D)=0$ by~\eqref{eq:anti}. Hence acceptance is impossible and~\eqref{eq:anti} gives $P(x_D)=1$. Soundness now excludes rejection.
\end{proof}
We call~\eqref{eq:anti} the \emph{anti-acceptance relation}. Its realization is the construction problem; the lemma supplies the common exclusion argument. An arbitrary sound evaluator can return $\bot$ even when the information determines the answer. To infer semantic separation from nondecision, one needs a mechanism that records all consequences of the background.

For a consistent classical theory $\Gamma$, define its deductive status on sentences by
\begin{equation}\label{eq:deductive-status}
 D_\Gamma(A)=
 \begin{cases}
 1,&\Gamma\vdash A,\\
 0,&\Gamma\vdash\neg A,\\
 \bot,&\text{otherwise}.
 \end{cases}
\end{equation}
Proposition~\ref{prop:relative} says that $D_\Gamma(A)=\bot$ exactly when the background has models with both truth values of $A$. This uses completeness of the deductive calculus, not completeness of the theory. In the arithmetic application of Lemma~\ref{lem:anti}, the objects are sentences, $V$ is the identity, and $P$ is standard truth; opposite models follow subsequently from Proposition~\ref{prop:relative}. In the finite application, the objects are complete formulas, $V$ is a clause observation, and $P$ is satisfiability.

\subsection{Two forms of reference}
The two constructions use the following forms of reference.

\emph{Encoded reference} starts with a code of an expression and a predicate on codes. A diagonal construction closes the reference by producing $G$ with $T\vdash G\leftrightarrow\psi(\code G)$. The construction, including the chosen coding and substitution operations, is part of the data.

\emph{Witness-dependent feedback} starts with two types of objects: sources $x\in\mathcal D$ and witnesses $w\in\mathcal W$, related by $E(x,w)$. If $x$ has a unique witness, write it as $\sigma(x)$. A modification is then specified by
\begin{equation}\label{eq:feedback}
 x\longmapsto\sigma(x)\longmapsto B(\sigma(x),\theta)
 \longmapsto K(x,\sigma(x),\theta),
\end{equation}
where $\theta$ records the permitted location or support of the modification. The operation is tied to the witness of that source, rather than an independently chosen witness. A feedback construction must separately establish what happens to all witnesses and what context is retained.

For solution blocking, \eqref{eq:feedback} is the semantic dependency called self-reference in~\cite{Fang}: the unique satisfying assignment determines a constraint that is inserted back into its mother formula. We use this specific dependency, rather than taking arbitrary dependence on an input property as a definition of self-reference.

For each mechanism, the proof must establish the effect of the negative operation and the preservation of the context. These are the steps that yield~\eqref{eq:split}.

\section{G\"odelian Self-Reference and Relative Incompleteness}\label{sec:godel}
\subsection{The idea before the coding}
In the framework of Section~\ref{sec:framework}, an arithmetic theory $T$ is the context, the models of $T$ are its realizations, and the target is the truth of a sentence in those models. To establish relative incompleteness, we need a sentence whose truth varies across models of the same $T$. G\"odel's construction produces such a target for each theory in the class considered below.

Consider a sentence intended to say: ``This sentence has no proof in $T$.'' Suppose that $T$ can verify concrete proofs and that the intended sentence has been constructed correctly. If $T$ proved it, the concrete proof would also let $T$ establish that it has a proof. The sentence asserts the opposite, so consistency rules out such a proof. The sentence then states a true fact about the absence of a proof. If $T$ is sound in the standard natural numbers, it cannot prove the sentence's negation either.

Two tasks remain before this becomes a mathematical argument. Arithmetic must express what a proof is, and the phrase ``this sentence'' must be replaced by a precise reference to the sentence being constructed. Coding makes expressions and proofs accessible as natural numbers. The diagonal lemma supplies the reference by an explicit substitution. We first carry out that construction, then prove independence, and finally use completeness to obtain opposite models of the retained theory. The assumption of soundness means that every theorem of $T$ is true in the standard natural numbers; it is stronger than consistency and is sufficient for the argument used here.

\subsection{Coding and the fixed-point construction}
Fix a standard effective coding of arithmetic expressions. If $G$ is a sentence, $\code G$ denotes its code, or the numeral for that code when it occurs inside an arithmetic formula. Substitution of a numeral into a formula is primitive recursive. Let $\operatorname{diag}(u)$ be the code obtained by substituting the numeral for $u$ into the designated free variable of the formula coded by $u$, with a fixed default value on malformed inputs. Its graph has an arithmetic representation $\Diag(u,v)$ whose totality and functionality are provable in $\PA$.

\begin{lemma}[Diagonal lemma]\label{lem:diagonal}
For each arithmetic formula $\psi(z)$ with one free variable, there is a sentence $G$ such that
\begin{equation}\label{eq:fixed}
 \PA\vdash G\leftrightarrow\psi(\code G).
\end{equation}
\end{lemma}
\begin{proof}
Let $A(u)$ be $\exists v(\Diag(u,v)\land\psi(v))$, let $a=\code{A(u)}$, and set $G=A(\bar a)$. By the definition of substitution, $\operatorname{diag}(a)=\code G$. Peano arithmetic verifies this concrete computation and the uniqueness of the output, so it proves $A(\bar a)\leftrightarrow\psi(\code G)$.
\end{proof}
This is the usual diagonal construction; see~\cite[Section 5.3]{Li} and~\cite{BBJ}. The formula $A$ is written before its code is substituted. No sentence is assumed to be available before it is constructed.

Let $T$ be a computably enumerable extension of $\PA$. One can use finite derivation records with enumeration-stage certificates for the axioms of $T$ to obtain a primitive recursive proof-checking relation. Its existential closure gives a standard predicate $\Prov_T(z)$. Applying Lemma~\ref{lem:diagonal} to $\neg\Prov_T(z)$ yields
\begin{equation}\label{eq:godel}
 T\vdash G_T\leftrightarrow\neg\Prov_T(\code{G_T}).
\end{equation}
The relation in~\eqref{eq:godel} is provability in $T$, not truth in an arbitrary model. This distinction is essential to its consistent interpretation.

\subsection{Relative incompleteness over fixed theories}
An independent sentence leaves two consistent extensions of the theory: one obtained by adjoining the sentence and one by adjoining its negation. Completeness turns these extensions into models. Thus the opposite possibilities are models of a common theory; they do not represent changing truth values within the standard natural numbers.

\begin{theorem}[Independence as relative incompleteness]\label{thm:independence}
Let $T$ be a consistent theory extending $\PA$, and let $G$ be a sentence. The following are equivalent:
\begin{enumerate}
\item $T\nvdash G$ and $T\nvdash\neg G$;
\item there are countable models $\mathcal M_+,\mathcal M_-$ of $T$ with
\[
 \mathcal M_+\models G,
 \qquad \mathcal M_-\models\neg G.
\]
\end{enumerate}
Thus $G$ is independent of $T$ exactly when $T$ is incomplete relative to $G$. In the semantic formulation of Proposition~\ref{prop:relative}, the two models witness separation over the context $T$ for the predicate $P_G(\mathcal M)=\one[\mathcal M\models G]$.
\end{theorem}
\begin{proof}
If $T\nvdash\neg G$, the deduction theorem implies that $T\cup\{G\}$ is consistent. Similarly, $T\cup\{\neg G\}$ is consistent when $T\nvdash G$. Completeness and downward L\"owenheim--Skolem give countable models of these theories. Conversely, soundness of first-order deduction rules out a proof of either $G$ or $\neg G$ from $T$ when the two models exist.
\end{proof}

\begin{theorem}[G\"odelian relative incompleteness]\label{thm:godel}
Let $\mathcal T$ be the class of computably enumerable extensions of $\PA$ sound in the standard natural numbers. For every $T\in\mathcal T$, the construction in~\eqref{eq:godel} gives a sentence $G_T$ such that $T$ is incomplete relative to $G_T$. Thus
\begin{equation}\label{eq:all-theories}
 \forall T\in\mathcal T\ \exists G_T:
 \quad T\nvdash G_T\quad\text{and}\quad T\nvdash\neg G_T.
\end{equation}
\end{theorem}
\begin{proof}
Use $D_T$ from~\eqref{eq:deductive-status}, with sentences as objects and $P(A)=\one[\N\models A]$. Soundness of $T$ makes this evaluator sound. The standard proof predicate satisfies $\N\models\Prov_T(\code A)$ exactly when $T\vdash A$. Since~\eqref{eq:godel} holds in $\N$,
\[
 P(G_T)=1-\one[T\vdash G_T]=1-a_{D_T}(G_T).
\]
Lemma~\ref{lem:anti} gives $D_T(G_T)=\bot$ and $\N\models G_T$. Theorem~\ref{thm:independence} supplies two models of $T$ in which $G_T$ has opposite truth values. They witness the relative incompleteness of $T$ for this target.
\end{proof}
The theory $T$ was arbitrary in the stated class. The construction does not merely identify a sentence omitted by one selected theory: it uses each theory's own proof predicate to construct a sentence that theory cannot decide. This is the universal guarantee in~\eqref{eq:all-theories}. The soundness assumption is a convenient sufficient hypothesis for the particular G\"odel sentence used here. Under consistency alone one may instead use a Rosser sentence~\cite{Rosser}. The present construction retains~\eqref{eq:godel} so that the reference and its negation remain explicit.

Both models in Theorem~\ref{thm:godel} satisfy every theorem of $T$. They are not elementarily equivalent: $G_T$ itself distinguishes them. In the model satisfying $\neg G_T$, the predicate $\Prov_T(\code{G_T})$ holds, although there is no standard finite proof. Such a model is nonstandard and may contain nonstandard purported proof codes. Its existence does not assert that a standard proof has been found.

The construction provides an independent sentence and, through completeness, models with opposite values. It does not specify an effective transformation from an arbitrary model of $T$ to a model with the opposite value of $G_T$. The separation is an existence statement over a common theory. Section~\ref{sec:blocking} applies the same criterion with the observed clauses as the retained information. There the two realizations are compatible completions, and the target that separates them is satisfiability.

\section{Finite Solution Blocking and Relative Incompleteness}\label{sec:blocking}
We first characterize when a single unobserved clause can be replaced to turn a satisfiable formula into an unsatisfiable one while preserving the observation. We then use this construction to establish relative incompleteness and determine exactly which observations suffice to decide satisfiability.

\subsection{The exact condition for one-clause blocking}
Example~\ref{ex:small} uses a unique global solution. More generally, a clause can exclude many solutions at once if they all agree on its variables. The relevant solutions are those remaining after the original clause is deleted. We begin with this exact condition.

Fix variables $x_1,\ldots,x_N$ and indexed supports $H=(J_1,\ldots,J_M)$, each nonempty. Every clause uses each variable of its support once, with either sign. An indexed formula $F=(C_1,\ldots,C_M)$ may have repeated clauses. Write $\Sol(F)$ for its solutions, $D_d=C_d$, and $Q_d=F-D_d$ for deletion of the indexed occurrence. The position is redundant when
\begin{equation}\label{eq:red}
 \Sol(Q_d)=\Sol(F).
\end{equation}
For $t\in\{0,1\}^{J}$ define
\begin{equation}\label{eq:blockclause}
 B_{t,J}=\bigvee_{i\in J}\ell_i,
 \qquad \ell_i=\begin{cases}x_i,&t_i=0,\\\neg x_i,&t_i=1.\end{cases}
\end{equation}
For a full assignment $s$, $B_{s,J}$ abbreviates $B_{s|_J,J}$. Its falsifying set is exactly $\{a:a|_J=t\}$. Equivalently, on Boolean integer variables,
\begin{equation}\label{eq:polynomial}
 B_{t,J}(x)=1\quad\Longleftrightarrow\quad
 \sum_{i\in J}((1-t_i)x_i+t_i(1-x_i))\ge1.
\end{equation}

\begin{theorem}[Projection criterion]\label{thm:projection}
Let $F$ be satisfiable and let $t\in\{0,1\}^{J_d}$. The following are equivalent:
\begin{enumerate}
\item $Q_d\land B_{t,J_d}$ is unsatisfiable;
\item $\pi_{J_d}(\Sol(Q_d))=\{t\}$;
\item $d$ is redundant in $F$ and $\pi_{J_d}(\Sol(F))=\{t\}$.
\end{enumerate}
In particular, a support-preserving replacement at $d$ makes $F$ unsatisfiable if and only if the retained solution set has a singleton projection on $J_d$.
\end{theorem}
\begin{proof}
The solution set after replacement is
\[
 \Sol(Q_d\land B_{t,J_d})
 =\{a\in\Sol(Q_d):a|_{J_d}\ne t\}.
\]
Since $F$ is satisfiable, $Q_d$ is satisfiable, proving the equivalence of the first two conditions. The third implies the second by redundancy. Conversely, under the second condition, choose a solution of $F$. Its restriction is $t$, so $t$ satisfies the original clause $C_d$. Every solution of $Q_d$ has that restriction and therefore satisfies $C_d$. Thus $d$ is redundant, and the solution sets, including their projections, agree. Every clause on $J_d$ is $B_{t,J_d}$ for exactly one $t$, proving the last assertion.
\end{proof}

Define the set of blocking positions by
\begin{equation}\label{eq:blockpositions}
 \mathcal B(F)=\{d:|\pi_{J_d}(\Sol(F-D_d))|=1\}.
\end{equation}
For each $d\in\mathcal B(F)$, let $t_d$ be the unique projection and let $F^{(d)}$ replace $C_d$ by $B_{t_d,J_d}$. The replacement preserves the indexed unsigned incidence graph, including every clause width. It differs from $F$ exactly at $d$, because $F$ is satisfiable and $F^{(d)}$ is not.

\begin{corollary}[Unique-solution blocking]\label{thm:block}
If $\Sol(F)=\{s\}$, then $\mathcal B(F)$ is exactly the set $\mathcal D(F)$ of redundant positions. For every such position,
\[
 F^{(d)}=(F-D_d)\land B_{s,J_d}
\]
is unsatisfiable with the same indexed supports as $F$.
\end{corollary}
\begin{proof}
Every projection of $\{s\}$ is a singleton, so Theorem~\ref{thm:projection} reduces the criterion to redundancy.
\end{proof}
This recovers the deterministic blocking step in~\cite{Fang} and proves its converse within the class of single-position, support-preserving replacements. Global uniqueness is sufficient but not necessary: on variables $x,y,z$, the formula $(x)\land(y)\land(x\lor y)$ has two solutions, differing in $z$, and replacing its third clause by $\neg x\lor\neg y$ excludes both.

\subsection{Diagonal reversal under an edit restriction}
Let $\mathcal X_H$ be all formulas with indexed supports $H$, and set $P(G)=\one[\Sol(G)\ne\varnothing]$. Let $V_S(G)$ record the clauses at positions $S$ and the public support system. Write $\mathcal N_S(F)$ for the formulas in $\mathcal X_H$ obtainable from $F$ by changing at most one position outside $S$, including $F$ itself. All its members have observation $V_S(F)$.

\begin{theorem}[Restricted diagonal realization]\label{thm:implementation}
For satisfiable $F$, let $\mathcal H$ be the set of all functions $h:V_S(\mathcal X_H)\to\{0,1\}$. The following are equivalent:
\begin{enumerate}
\item There is a map $C:\mathcal H\to\mathcal N_S(F)$ such that, for every $h$,
\begin{equation}\label{eq:restricted-diagonal}
 P(C(h))=1-h(V_S(C(h))).
\end{equation}
\item $\mathcal N_S(F)$ contains an unsatisfiable formula.
\item $\mathcal B(F)\nsubseteq S$.
\end{enumerate}
When these conditions hold, one fixed $d\in\mathcal B(F)\setminus S$ gives the construction
\begin{equation}\label{eq:selector}
 C(h)=\begin{cases}F^{(d)},&h(V_S(F))=1,\\F,&h(V_S(F))=0.\end{cases}
\end{equation}
\end{theorem}
\begin{proof}
Apply the first condition to the constant function $h\equiv1$ to obtain the second. An unsatisfiable member of $\mathcal N_S(F)$ differs from the satisfiable $F$ at exactly one position $d\notin S$. Theorem~\ref{thm:projection} gives $d\in\mathcal B(F)$, proving the third. Under the third condition, $P(F)=1$, $P(F^{(d)})=0$, and both branches of~\eqref{eq:selector} preserve $V_S(F)$. These identities prove the first condition.
\end{proof}
The functions $h$ in this theorem are arbitrary binary predictions, not assumed correct. For a sound partial evaluator $D$, substitute its acceptance indicator $h=a_D$ in~\eqref{eq:selector}. The resulting object satisfies~\eqref{eq:anti}; Lemma~\ref{lem:anti} gives $D(V_S(F))=\bot$. Once nondecision is established, the selector returns $F$. Thus the conclusion concerns the original fixed mother formula, even though the proof uses a conditional choice. This construction asserts existence without a bound on the cost of finding $d$, $t_d$, or an evaluator's status.

\begin{corollary}\label{cor:views}
Every observation $S$ with $\mathcal B(F)\nsubseteq S$ admits an unsatisfiable completion agreeing with the satisfiable $F$. If $F$ is uniquely satisfiable, this includes every $S$ omitting any redundant position, and hence every $|S|<|\mathcal D(F)|$.
\end{corollary}
\begin{proof}
Choose $d\in\mathcal B(F)\setminus S$ and use $F^{(d)}$.
\end{proof}
The necessity in Theorem~\ref{thm:implementation} concerns one-position replacements. Observations covering $\mathcal B(F)$ may still admit unsatisfiable completions differing at several positions.

\subsection{Relative incompleteness under local observations}\label{sec:local-incompleteness}
The observation theory is the context, the compatible completions are its realizations, and the target is the satisfiability of the complete formula. There are two kinds of Boolean variables to keep separate. The variables $x_i$ describe a candidate solution to a CNF. The variables introduced below describe which literals occur in the CNF. An observer knows some of these input signs and asks whether the full input admits a solution. Encoding that question as one propositional assertion lets ordinary logical consequence express exactly what the observation determines.

Fix an indexed support system $H=(J_1,\ldots,J_M)$ on $[N]$, with each $J_j$ nonempty. For each $i\in J_j$, introduce a propositional variable $z_{j,i}$: value $0$ encodes the literal $x_i$, and value $1$ encodes $\neg x_i$. A valuation $z$ therefore encodes a complete CNF $F_z$ with supports $H$.

For a fixed assignment $a\in\{0,1\}^N$, let
\[
 L_{a_i}(z_{j,i})=
 \begin{cases}z_{j,i},&a_i=0,\\ \neg z_{j,i},&a_i=1,\end{cases}
 \qquad
 E_{H,a}(z)=\bigwedge_{j=1}^{M}\bigvee_{i\in J_j}L_{a_i}(z_{j,i}).
\]
Thus $E_{H,a}(z)$ holds exactly when $a$ satisfies $F_z$. Define the finite propositional assertion
\begin{equation}\label{eq:finite-target}
 A_H(z)=\bigvee_{a\in\{0,1\}^N}E_{H,a}(z).
\end{equation}
Its truth value is the satisfiability of $F_z$. The disjunction gives a semantic encoding; no bound on its succinctness is needed.

Write $z^F$ for the sign encoding of $F$. For $S\subseteq[M]$, let
\begin{equation}\label{eq:observation-theory}
 \Gamma_S(F)=
 \{z_{j,i}:j\in S,\ z^F_{j,i}=1\}
 \cup\{\neg z_{j,i}:j\in S,\ z^F_{j,i}=0\},
\end{equation}
where $i$ ranges over $J_j$. Its models are exactly the encodings of formulas with supports $H$ agreeing with $F$ at the observed positions. These axioms describe the \emph{input formula}, rather than imposing its clauses as constraints on an assignment $x$. The support system is fixed and public. The observer is not given a promise that the unknown full formula is uniquely satisfiable; such a promise would already determine $A_H$.

\begin{theorem}[Relative incompleteness under local observations]\label{thm:finite}
For satisfiable $F$ with supports $H$, every $S$ satisfying $\mathcal B(F)\nsubseteq S$ obeys
\begin{equation}\label{eq:finite-independent}
 \Gamma_S(F)\nvdash A_H,\qquad \Gamma_S(F)\nvdash\neg A_H.
\end{equation}
These statements persist after adjoining all logical consequences of the observed facts. The full description $\Gamma_{[M]}(F)$ proves $A_H$. For uniquely satisfiable $F$, the hypothesis is exactly $\mathcal D(F)\nsubseteq S$.
\end{theorem}
\begin{proof}
For any attainable observation $y$, let $\Gamma_y$ be its unit sign facts and let $D(y)$ be their deductive status on $A_H$. Soundness of propositional deduction makes $D$ sound for satisfiability. The selector following Theorem~\ref{thm:implementation} and Lemma~\ref{lem:anti} force $D(V_S(F))=\bot$, giving~\eqref{eq:finite-independent}. Explicitly, $z^F$ and $z^{F^{(d)}}$, for $d\in\mathcal B(F)\setminus S$, are opposite models of $\Gamma_S(F)$. Proposition~\ref{prop:closure} preserves them under consequences. The full description has the sole valuation $z^F$, which satisfies $A_H$, so propositional completeness proves the full-information claim. Corollary~\ref{thm:block} gives the unique-solution specialization.
\end{proof}

\subsection{All opposite completions and indispensable positions}
Section~\ref{sec:local-incompleteness} gives a sufficient condition for relative incompleteness based on a single unobserved edit. To characterize exactly when a local observation determines satisfiability, we consider all compatible unsatisfiable completions. For satisfiable $F\in\mathcal X_H$, define
\[
 \Delta(F,G)=\{j:C_j(F)\ne C_j(G)\},\qquad
 \mathcal E_F=\{\Delta(F,G):G\in\mathcal X_H,\ \Sol(G)=\varnothing\}.
\]
These difference sets are nonempty. The family may itself be empty; then all formulas with supports $H$ are satisfiable.

\begin{theorem}[Sufficient observations]\label{thm:observations}
For satisfiable $F$, an observation $S$ proves $A_H$ if and only if it intersects every member of $\mathcal E_F$:
\begin{equation}\label{eq:hitting}
 \Gamma_S(F)\vdash A_H
 \quad\Longleftrightarrow\quad
 \forall E\in\mathcal E_F,\ S\cap E\ne\varnothing.
\end{equation}
Otherwise $A_H$ is independent of $\Gamma_S(F)$. The condition on the right is vacuous when $\mathcal E_F$ is empty.
\end{theorem}
\begin{proof}
An unsatisfiable $G$ agrees with the observation exactly when $S\cap\Delta(F,G)=\varnothing$. Thus the right side says that every model of the observed sign facts encodes a satisfiable formula. Propositional soundness and completeness give~\eqref{eq:hitting}. If a disjoint difference set remains, its formula and $F$ provide opposite models. Since $F$ always satisfies the observation, $\Gamma_S(F)$ never proves $\neg A_H$.
\end{proof}

Call $S$ \emph{sufficient at $F$} if it satisfies~\eqref{eq:hitting}, and call $d$ \emph{indispensable at $F$} if every sufficient $S$ contains $d$.
\begin{corollary}[Indispensable positions]\label{cor:indispensable}
For satisfiable $F$, the following are equivalent:
\[
 d\in\mathcal B(F),\qquad
 \{d\}\in\mathcal E_F,\qquad
 d\text{ is indispensable at }F.
\]
For uniquely satisfiable $F$, these are exactly the redundant positions.
\end{corollary}
\begin{proof}
The first equivalence is Theorem~\ref{thm:projection}. A singleton difference set forces every sufficient observation to contain $d$. Conversely, if $\{d\}\notin\mathcal E_F$, the set $[M]\setminus\{d\}$ intersects every difference set, since none is empty or contained in $\{d\}$. It is therefore sufficient without $d$. Apply Corollary~\ref{thm:block} for uniqueness.
\end{proof}
Redundancy concerns a clause's effect on the solution set; indispensability concerns knowledge of the input signs. A redundant clause can be deleted without losing any constraint on solutions, yet its unseen position could hold a clause excluding them all.

\begin{example}\label{ex:necessary}
Let $F=(x)\land(y)\land(x\lor y)$. Its sole solution is $(1,1)$ and its sole redundant position is $3$. Observing position $3$ alone is insufficient: $(\neg x)\land(\neg y)\land(x\lor y)$ is an unsatisfiable completion. The inclusion-minimal difference sets are exactly $\{3\}$ and $\{1,2\}$. Indeed, the two unit clauses fix an assignment, and the binary clause causes unsatisfiability exactly when it excludes that assignment. If position $3$ changes, the difference set contains $\{3\}$; if it does not, both units must become negative, giving $\{1,2\}$. Both minimal sets occur. Thus $S$ is sufficient exactly when $3\in S$ and $S\cap\{1,2\}\ne\varnothing$.
\end{example}
The example separates indispensable information from sufficient information. It also shows why the single-position criterion cannot characterize every source of relative incompleteness.

\section{Diagonal Realizations and Arithmetization}\label{sec:comparison}
The preceding results share the anti-acceptance relation but realize it in different ways. We first make its connection to the set-theoretic diagonal pattern explicit. We then describe blocking in arithmetic and distinguish that description from the construction responsible for independence.

\subsection{An evaluation-map realization}
The elementary Lawvere--Yanofsky pattern takes $e:A\times A\to\{0,1\}$ and forms $g(a)=1-e(a,a)$. For every $b$, the values $g(b)$ and $e(b,b)$ differ; hence $g$ is not any column $e(-,b)$~\cite{Lawvere,Yanofsky}.

\begin{proposition}[Evaluation map for anti-acceptance]\label{prop:evaluation}
Let a nonempty set $A$ index acceptance predicates $\operatorname{Acc}_a:\mathcal X\to\{0,1\}$ and objects $x_a$ satisfying $P(x_a)=1-\operatorname{Acc}_a(x_a)$. Then
\[
 e(a,b)=\operatorname{Acc}_b(x_a),\qquad g(a)=P(x_a)
\]
instantiate the binary diagonal pattern, and $g$ is not represented by any column of $e$. If the predicates are acceptance indicators of sound partial evaluators, each corresponding evaluator leaves $x_a$ undecided.
\end{proposition}
\begin{proof}
The hypothesis gives $g(a)=1-e(a,a)$. Evaluating at $b$ excludes column $b$. The final assertion is Lemma~\ref{lem:anti}.
\end{proof}
For arithmetic, index sound computably enumerable presentations $T_a$ extending $\PA$, take $x_a=G_{T_a}$, and put $\operatorname{Acc}_b(A)=\one[T_b\vdash A]$. Standard truth is $P$. Theorem~\ref{thm:godel} verifies the required relation via diagonal substitution; the index set of sound presentations need not be effectively enumerable.

For fixed $F,S$ in Theorem~\ref{thm:implementation}, index any family of partial evaluators $D_a$ of $V_S$ and put $x_a=C(a_{D_a})$. Set $\operatorname{Acc}_b(G)=a_{D_b}(V_S(G))$. The selector realizes the same relation. All $x_a$ have the observation $y=V_S(F)$, so in this instance $e(a,b)=a_{D_b}(y)$ is independent of $a$. Every column is constant. This exact embedding captures the common diagonal relation for opposite completions compatible with the same observation. Solution blocking gives that relation a concrete realization under clause-edit constraints.

Indeed, for a fixed observation $y$ and without an edit restriction, compatible completions attain both target values if and only if every binary predictor $h$ can be opposed by some compatible completion $x_h$ with $P(x_h)=1-h(y)$. One direction selects the opposite value; the other uses the constant predictors zero and one. Theorem~\ref{thm:implementation} adds the structural question omitted by that equivalence: when can this choice be realized by one unobserved clause replacement on fixed supports? The answer is the projection criterion, and Corollary~\ref{cor:indispensable} identifies its observational consequence. The contribution is a constrained finite realization of the set-theoretic pattern, with an exact criterion linking the permitted edit to the solution structure.

\subsection{The two guarantees and their mechanisms}
\begin{theorem}[Two realizations of relative incompleteness]\label{thm:comparison}
The following constructions realize~\eqref{eq:anti} and establish the guarantees in~\eqref{eq:class-guarantee}.
\begin{enumerate}
\item For every sound computably enumerable extension $T$ of $\PA$, diagonal substitution constructs $G_T$ with $D_T(G_T)=\bot$. The theory $T$ is incomplete relative to $G_T$, whose truth has opposite values in two models of $T$.
\item Fix a satisfiable $F$ and an observation $S$ with $\mathcal B(F)\nsubseteq S$. The selector~\eqref{eq:selector}, applied to the acceptance indicator of deductive status, shows that $\Gamma_S(F)$ is incomplete relative to $A_H$. For uniquely satisfiable $F$, this covers every $S$ omitting a redundant position.
\end{enumerate}
Both separations persist under deductive closure. In the second case, a support-preserving realization using at most one unobserved edit exists exactly under the stated condition on $\mathcal B(F)$.
\end{theorem}
\begin{proof}
Theorem~\ref{thm:godel} gives two models of $T$ assigning opposite truth values to $G_T$. Theorems~\ref{thm:implementation} and~\ref{thm:finite} give opposite completions, represented by a satisfiable formula and its unsatisfiable companion. Proposition~\ref{prop:relative} identifies these pairs as witnesses to relative incompleteness in their respective contexts. Proposition~\ref{prop:closure} gives preservation under consequences.
\end{proof}
In arithmetic, the sentence refers to the proof relation for itself. In finite blocking, the solution set determines an excluding clause; a separate choice between the source and its companion opposes the evaluator's acceptance. This latter choice links the solution-dependent modification to the common diagonal argument; encoded self-reference is supplied separately by the tagged arithmetic construction below. Table~\ref{tab:comparison} records the distinction.

\begin{table}[ht]
\centering\small
\renewcommand{\arraystretch}{1.15}
\begin{tabularx}{\textwidth}{@{}>{\raggedright\arraybackslash}p{0.22\textwidth}>{\raggedright\arraybackslash}X>{\raggedright\arraybackslash}X@{}}
\toprule
Component & Arithmetic realization & Finite realization\\
\midrule
Internal dependency & Sentence's own proof predicate & Source's solution projection\\
Construction & Diagonal substitution & Blocking and conditional selection\\
Retained background & Fixed theory $T$ & Fixed observation $V_S(F)$\\
Target & $G_T$, varying with $T$ & Fixed satisfiability assertion $A_H$\\
Realization condition & Coding and a sound effective theory & Singleton projection at an unseen position\\
Opposite models & Models of $T+G_T$ and $T+\neg G_T$ & Sign encodings of mother and companion\\
\bottomrule
\end{tabularx}
\caption{Two implementations of the common anti-acceptance argument.}
\label{tab:comparison}
\end{table}

\subsection{Arithmetic predicates for solution blocking}
A blocking assertion says that the mother formula has one solution, removing the selected clause preserves its solutions, and the replacement excludes that solution. Each condition can be checked by a finite procedure. Coding the procedure turns this description into an arithmetic statement, with the mother and companion kept as distinct objects.

Choose a primitive recursive coding of indexed CNFs, supports, and assignments. The declared number of variables $n(f)$ and the number of clauses $m(f)$ are part of a valid formula code $f$. A number $a<2^{n(f)}$ encodes an assignment through its binary digits. Let $\Sat(f,a)$ mean that $f$ is a valid CNF code, $a$ is in range, and the encoded assignment satisfies every clause. Validation, bit extraction, and evaluation are primitive recursive.

We use symbols for these functions in a conservative definitional extension of $\PA$. They can be eliminated using their graph formulas. In particular, exponentiation and the finite sequence operations below have PA-provable totality. The predicates are chosen from their direct finite algorithms, so their elementary correctness identities are provable by induction in $\PA$.

Define
\begin{align}
 \Unique(f,a)\;:\!\Longleftrightarrow\;&
 \Sat(f,a)\land
 \forall b<2^{n(f)}\bigl(\Sat(f,b)\to b=a\bigr),\label{eq:unique}\\
 \Red(f,d)\;:\!\Longleftrightarrow\;&
 1\le d\le m(f)\land
 \forall b<2^{n(f)}
 \bigl(\Sat(f,b)\leftrightarrow\Sat(\operatorname{Del}(f,d),b)\bigr).
 \label{eq:arithred}
\end{align}
Deletion preserves the declared variable universe. Let $\Rep(f,d,a,J)$ be the code obtained by replacing position $d$ with the clause~\eqref{eq:blockclause} specified by $a$ and $J$. Invalid inputs receive a fixed default output. Write $\operatorname{Adm}(f,d,J)$ for the primitive recursive condition that $f$ is an indexed CNF with nonempty clauses on distinct variables, $d$ is a valid position, and $J=\supp(D_d)$. If exact width is required, include that check as well.

\begin{definition}
The arithmetic \emph{blocking relation} is
\begin{equation}\label{eq:blockarith}
\begin{split}
 \operatorname{Block}(f,g)\;:\!\Longleftrightarrow\;
 \exists a,d,J\,[\,&\operatorname{Adm}(f,d,J)\land\Unique(f,a)\\
 &\land\Red(f,d)\land g=\Rep(f,d,a,J)\,].
\end{split}
\end{equation}
Set $\operatorname{Unsat}(g):=\operatorname{ValidCNF}(g)\land
\forall b<2^{n(g)}\neg\Sat(g,b)$.
\end{definition}

\begin{proposition}[Arithmetic correctness]\label{prop:arithmetic}
For the encodings just specified,
\[
 \PA\vdash\forall f\,\forall g\,
 \bigl(\operatorname{Block}(f,g)\to\operatorname{Unsat}(g)\bigr).
\]
Peano arithmetic also proves equality of the declared variable counts and clause counts, and equality of the indexed unsigned incidence graphs, for every blocking pair.
\end{proposition}
\begin{proof}
Work in $\PA$ and take witnesses $a,d,J$ to~\eqref{eq:blockarith}. If $\Sat(g,b)$ holds, correctness of replacement implies $\Sat(\operatorname{Del}(f,d),b)$. Redundancy gives $\Sat(f,b)$, and uniqueness gives $b=a$. Correctness of the blocking clause says that any satisfying assignment of that clause differs from $a$ on at least one position in $J$. This contradicts $b=a$. Validity of $g$ and preservation of its declared parameters follow from admissibility. The graph identity follows because replacement retains the position and support. Each step is an identity or a bounded universal implication for the primitive recursive operations, verified by the corresponding finite-list inductions in $\PA$.
\end{proof}
Thus the informal assertion that a unique solution is forbidden has a precise arithmetic interpretation. It relates the mother code $f$ to the companion code $g$. The assignment is unique for $f$; it is not a satisfying assignment of $g$. We now use these predicates to construct an arithmetic sentence that asserts the validity of the blocking data encoded in the sentence itself.

\subsection{A self-referential sentence for solution blocking}
The arithmetic sentence $G_p$ has the following natural-language reading:
\begin{quote}
\emph{The data carried in my own code specify a valid unique-solution blocking construction.}
\end{quote}
A syntactic tag stores those data, and a parsing function reads the tag. Diagonal substitution makes the argument of that parsing function the code of the completed sentence itself. The resulting fixed point expresses self-reference at the coding level; its mathematical content remains the assertion about the stored data.

The relation~\eqref{eq:blockarith} can itself be placed in a diagonal sentence. Let $p$ code a triple $(f,d,J)$, and define
\begin{equation}\label{eq:rp}
 R(p):=\operatorname{Triple}(p)\land\operatorname{Adm}(f,d,J)
 \land\Red(f,d)\land\exists a\,\Unique(f,a),
\end{equation}
where the components are obtained by the fixed projection functions. The companion code is then $\Rep(f,d,a,J)$; the unique $a$ determines it, and Proposition~\ref{prop:arithmetic} proves its unsatisfiability. Thus $R$ asserts the validity of the blocking construction: the mother has a unique solution, and the companion excludes it.

Choose a syntax convention in which formulas may start with a tag $\bar p=\bar p$ as the left conjunct. Let $\Payload(z)$ extract the numeral from this leading tag, using a default value on other expressions. This is a primitive recursive parsing function. We keep the syntactic tag in the written formula even though it is logically redundant.

\begin{proposition}[Tagged diagonal embedding]\label{prop:tagged}
For every natural number $p$, one can effectively construct an arithmetic sentence $G_p$ such that
\begin{align}
 \PA&\vdash G_p\leftrightarrow
 R\bigl(\Payload(\code{G_p})\bigr),\label{eq:tagfixed}\\
 \PA&\vdash\Payload(\code{G_p})=\bar p,\label{eq:extract}\\
 \PA&\vdash G_p\leftrightarrow R(\bar p).\label{eq:collapse}
\end{align}
\end{proposition}
\begin{proof}
With the coding of Section~\ref{sec:godel}, form
\[
 A_p(u):=(\bar p=\bar p)\land
 \exists v\bigl(\Diag(u,v)\land R(\Payload(v))\bigr).
\]
Let $a_p=\code{A_p(u)}$ and set $G_p=A_p(\bar a_p)$. The diagonal function sends $a_p$ to $\code{G_p}$. Its concrete computation and functionality, together with the true tag equation, give~\eqref{eq:tagfixed}. Substitution does not change the leading tag, so direct parsing gives~\eqref{eq:extract}. Substitution of this equality gives~\eqref{eq:collapse}. All abbreviations may be expanded before coding; the leading tag is retained by this convention.
\end{proof}
Equations~\eqref{eq:tagfixed}--\eqref{eq:collapse} have a direct reading. The first says that $G_p$ asserts the blocking property of the data read from its own code. The second identifies those data as $p$. The third therefore reduces the assertion to $R(\bar p)$. For each fixed valid blocking datum, finite verification proves $R(\bar p)$ in $\PA$, and hence proves $G_p$.

Independence depends on the predicate and reference mechanism used in the diagonal construction. In~\eqref{eq:godel}, the predicate applied to the sentence's code denies the existence of a proof of that very sentence in $T$. A hypothetical proof then supplies precisely the object whose existence the sentence denies. In~\eqref{eq:tagfixed}, the predicate instead reads a fixed datum and checks its blocking property. For valid data, a proof of $G_p$ verifies precisely the blocking property it asserts. The same tagged construction works for any arithmetic predicate in place of $R$; its content depends on that predicate. This is why the reference mechanism must be specified in addition to the fixed-point equivalence~\cite{HV,GHY}.

The finite independence proved in Theorem~\ref{thm:finite} belongs to a different pair: the observation theory $\Gamma_S(F)$ and the target $A_H$. It arises because the observation theory has both a satisfiable and an unsatisfiable completion. Verification from the full blocking data establishes the construction, while the observer's background $\Gamma_S(F)$ continues to admit both completions. A proof that specified information fails to determine an answer is compatible with that failure of determination.

The two displayed self-reference relations therefore have different roles. Equation~\eqref{eq:anti} concerns acceptance and drives the shared nondecision argument. Equation~\eqref{eq:tagfixed} concerns data extraction from a sentence's code and records the finite construction. The observation theorem uses $A_H$ as its target, and the tagged sentence describes the construction at the coding level.

\section{Logical Foundations of Hardness Hypotheses}\label{sec:discussion}

The preceding results show how a theory or a local observation can leave a target undecided. We now ask how this kind of information constraint can guide the formulation of computational hardness hypotheses.

\subsection{Meta-language environments and the self-recognition obstacle}

In the parable behind the Chinese idiom \emph{m\'ang r\'en m\=o xi\`ang}, each observer feels a different part of an elephant and tries to infer its overall shape. Suppose an observer can feel only one specified part. Two objects can feel identical at that part and have different overall shapes. Any judgment based solely on the permitted touch must be the same for both. To establish this limitation, we take the observation method itself as an object of study and compare the objects that can give rise to the same observation.

Li's principle of environments provides a framework for distinguishing these two kinds of reasoning~\cite[Secs.~10.1--10.2, Principle~10.1]{Li}. An object language provides the syntax, a model interprets it, and the meta-language environment is where we define the language and its models and prove results about them. An arithmetic derivation proceeds under the rules of a given theory. A proof that the theory leaves a sentence undecided examines the theory's axioms and provability relation. Similarly, a SAT decision rule reasons from the observed clauses, while a proof about its limits examines the complete formulas compatible with those clauses.

The \emph{self-recognition obstacle} discussed by Fang et al.~\cite[Sec.~6.1]{Fang} arises when a system uses its own methods to investigate the limits of those very methods. Such an investigation inherits the limits of the underlying background whenever it is confined to that background. The observer in the parable may recognize that the permitted touch leaves the overall shape open. If the observer uses that same touch to check whether a particular guess is correct, the check still cannot distinguish the two shapes. The meta-language analysis exposes the blind spot by comparing both possibilities and showing what the observation leaves unresolved.

Proposition~\ref{prop:determination} makes the underlying information limit precise. An admissible background fails to determine a binary target exactly when it admits two compatible realizations with opposite target values. Analytic, algebraic, and geometric methods all face this limit when their judgments depend only on the same background. Suppose a rule confined to that background were correct on both realizations. It would have to give different answers from identical information. This contradiction shows that a sound partial rule must leave the target undecided on that background.

Self-reference gives concrete witnesses to this limit in the two settings studied here. The G\"odel construction uses the theory's own provability predicate to obtain an independent sentence (Theorem~\ref{thm:godel}). Two models of the theory then disagree on that sentence. In the finite construction, the unique solution of the mother formula determines the blocking clause. For each observation that leaves a redundant clause position unseen, replacement at such a position preserves the observation and changes a satisfiable formula into an unsatisfiable one (Theorem~\ref{thm:finite}). The pair exposes a limitation shared by every rule that works from that observation. A statement about the limits of an entire class of methods is also the starting point for the complexity question.

\begin{table}[!b]
\centering\small
\setlength{\tabcolsep}{5pt}
\renewcommand{\arraystretch}{1.12}
\begin{tabularx}{\textwidth}{@{}>{\raggedright\arraybackslash}p{0.17\textwidth}*{3}{>{\raggedright\arraybackslash}X}@{}}
\toprule
 & \textbf{Arithmetic theories} & \textbf{Finite SAT observations} & \textbf{General algorithmic complexity}\\
\midrule
Internal method &
Derive a sentence using the theory's axioms and rules. &
Judge satisfiability from the prescribed observation. &
Run an algorithm in the specified computational model.\\
\addlinespace
Meta-level question &
Which sentences lie beyond the theory's power to decide? &
Which observations leave satisfiability unresolved for every rule confined to the observed information? &
What resource limits apply across the entire class of algorithms allowed by the model?\\
\addlinespace
Basis for the analysis &
Two models of the theory disagree on the target sentence. &
Two formulas agree on the observation and disagree on satisfiability. &
An identified structural obstacle motivates a resource claim about a specified class of algorithms.\\
\addlinespace
Status here &
The G\"odel construction gives an independent sentence under the hypotheses of Theorem~\ref{thm:godel}. &
Solution blocking gives opposite completions under the hypotheses of Theorem~\ref{thm:finite}. &
SRK-ESH states a worst-case time hypothesis for a specified promise family.\\
\bottomrule
\end{tabularx}
\caption{Internal methods and metamathematical questions.}
\label{tab:metamathematical-levels}
\end{table}

\subsection{Algorithms and complexity at two levels}

An algorithm is a mathematical method specified so that it can be carried out mechanically. To establish an unconditional upper bound, it suffices to construct one algorithm and prove that it solves the problem correctly within the stated resource bound. A general lower bound concerns every correct algorithm permitted by the specified computational model. It is metamathematical because it takes that entire class of algorithms as its object of study. Since algorithms are mechanized mathematical methods, studying their limits amounts to studying the limits of mathematics in its mechanized form.

An inquiry conducted solely through the methods under study inherits their blind spots. If proving an impossibility requires a distinction that those methods cannot make, then mathematics, working within those very methods, cannot establish that impossibility unconditionally. This is the self-recognition obstacle discussed in Section~6.1.

The finite SAT setting gives a concrete example. Suppose a procedure receives only the prescribed clause observation and is asked to verify the answer ``satisfiable'' for the actual input. If a satisfiable formula and an unsatisfiable formula give the same observation, the procedure has the same information in both cases, even though the proposed answer is correct for only one of them. At the metamathematical level, comparing the two compatible formulas establishes a limitation shared by every procedure confined to that observation.

Table~\ref{tab:metamathematical-levels} compares what a method does within each framework with the questions we can ask about the framework itself.

The first two settings show how a metamathematical result can identify a limitation shared by every method subject to a stated restriction. The third raises the question taken up in Section~\ref{sec:hardness-hypotheses}. How can a result of this kind guide the formulation of a hypothesis about the resource limits of an entire class of algorithms?

\subsection{From relative incompleteness to hardness hypotheses}\label{sec:hardness-hypotheses}

Proving $\mathrm{P}\ne\mathrm{NP}$ has long been a central challenge in computational complexity. First, as discussed above, algorithm design takes place at the mathematical level, while the study of algorithmic limits takes place at the metamathematical level. Second, despite decades of work, unconditional superlinear worst-case time lower bounds remain elusive for SAT and many other canonical $\mathrm{NP}$-complete problems in general-purpose deterministic models with no separate restriction on space. These two considerations lead us to think that $\mathrm{P}\ne\mathrm{NP}$ is quite likely to be independent of a standard mathematical theory such as ZFC. This suggests that the conventional order of inquiry may therefore have been reversed. A conjecture whose proof has long been a goal could itself serve as a working hypothesis from which to develop a theory of computational hardness. This shift makes the choice of a working hypothesis a methodological question in its own right.\footnote{Rapid advances in AI may be ushering in an era of industrial-scale mathematical proof production. As more of the work of finding and checking proofs becomes amenable to automation, formulating fruitful conjectures and establishing which hypotheses are well grounded may become an increasingly central task in mathematics and theoretical computer science.} How suitable is $\mathrm{P}\ne\mathrm{NP}$ as a starting point, and what grounds should guide our choice of hardness hypotheses more generally?

Two familiar grounds for adopting $\mathrm{P}\ne\mathrm{NP}$ are polynomial-time reductions and the experience of algorithm design. Reductions show that a polynomial-time algorithm for any $\mathrm{NP}$-complete problem would yield polynomial-time algorithms for all problems in $\mathrm{NP}$. The continued failure to find such an algorithm lends the conjecture empirical support. The logical foundation sought here would identify a structural feature of an $\mathrm{NP}$-complete problem that poses a common obstacle to every candidate polynomial-time algorithm. The arguments from reductions and algorithm design leave that obstacle unidentified.

The histories of the Exponential Time Hypothesis (ETH) and the 3SUM hypothesis illustrate the limits of this evidence~\cite{ImpagliazzoPaturi, VassilevskaWilliams}. ETH posited that deciding $3$-SAT on $n$ variables requires exponential time in the worst case. OpenAI refuted it by giving a deterministic algorithm that counts the satisfying assignments of a fixed-width $k$-CNF formula in $2^{o(n)}$ time up to polynomial factors. Testing whether the count is positive decides $3$-SAT within the same bound~\cite[Theorem~1.1]{OpenAIETH}. The 3SUM hypothesis ruled out an $O(n^{2-\varepsilon})$-time algorithm for any fixed $\varepsilon>0$. Alman and Vassilevska Williams refuted it for polynomially bounded integers with a deterministic $O(n^{1.9992})$-time algorithm~\cite{AlmanVassilevskaWilliams}. In both cases, a new algorithm crossed the time threshold that earlier work had given researchers reason to conjecture.

A metamathematical foundation for a hardness hypothesis begins with a structural obstacle shared by every method subject to an explicit restriction. Relative incompleteness supplies such an obstacle. In arithmetic, two models of the same theory can disagree on a sentence. In the finite setting, two complete formulas can agree on all observed clauses and differ in satisfiability. Any method confined to the prescribed information has the same basis for judgment in both cases. Since the correct answers differ, that information cannot determine the target.

Fang et al. propose a dual-evidence criterion for hardness hypotheses~\cite[Sec.~7.2]{Fang}. In the terms used here, a well-motivated hardness hypothesis should draw on both mathematical and metamathematical evidence. At the mathematical level, the analysis of known algorithms shows which speedups they achieve and where their techniques fall short. At the metamathematical level, a logical argument identifies a structural obstacle shared by all methods subject to an explicit restriction and explains its relevance to the conjectured resource bound. Both kinds of evidence should bear on the same instance family, computational model, and resource scale.

The Self-Referential $K$-SAT Exhaustive-Search Hypothesis (SRK-ESH) illustrates how these two kinds of evidence motivate a concrete resource claim. If an observation of a uniquely satisfiable mother formula omits a redundant clause position, blocking the unique solution at that position yields an unsatisfiable companion that agrees with the observation. Relative incompleteness thus identifies a distinction that every correct solver on the promise family must make and that the prescribed local information leaves unresolved. On the algorithmic side, the strategies examined by Fang et al. yield no established fixed proportional saving in the exponent on the same promise family when given the complete formula. Against the exhaustive-search benchmark of $2^N$ assignment tests, SRK-ESH asserts that, for every fixed $0<\delta<1$, no uniform deterministic algorithm decides the promise family in worst-case time $O^*(2^{(1-\delta)N})$, where $N$ is the number of variables and $O^*$ suppresses factors polynomial in the encoded input length~\cite[Sec.~7.2]{Fang}.

\section{Conclusion}\label{sec:conclusion}
Relative incompleteness gives a common characterization of G\"odelian incompleteness and the limits of local observations in finite SAT. In arithmetic, diagonalization yields a sentence on which two models of the same theory disagree. In finite SAT, solution blocking yields two formulas that agree on the observed clauses but differ in satisfiability. The common relation between truth and acceptance explains why a sound partial evaluator must leave the corresponding question undecided. For a satisfiable formula, a single support-preserving replacement can make it unsatisfiable exactly when the solutions remaining after deletion all agree on the support of the replaced clause. For a uniquely satisfiable mother formula, the eligible positions are exactly its redundant positions, and these are precisely the positions that every sufficient clause observation must include. The evaluation map connects both constructions to the established set-theoretic diagonal pattern. The arithmetic blocking relation and tagged fixed point describe the finite construction at the coding level.

A background and its deductive closure admit the same models, so sound reasoning preserves any opposite realizations compatible with it. This information perspective is related to Chaitin's analysis of formal systems~\cite{Chaitin}. Which judgments the available information supports is itself a claim that can be proved at the meta-language level. In the finite setting, relative incompleteness shows which prescribed observations leave satisfiability undecided. This obstruction motivates the SRK-ESH, whose quantitative claim concerns the worst-case time needed to decide the specified promise family.


\begin{thebibliography}{FLL+26}
\small
\setlength{\itemsep}{2pt}
\bibitem[AVW26]{AlmanVassilevskaWilliams}
Josh Alman and Virginia Vassilevska Williams.
Truly subquadratic 3SUM and truly subcubic APSP via triangles in sparse lopsided graphs.
\href{https://arxiv.org/abs/2610.06783}{arXiv:2610.06783}, 2026.

\bibitem[BBJ07]{BBJ}
George S. Boolos, John P. Burgess, and Richard C. Jeffrey.
\emph{Computability and Logic}. 5th edition. Cambridge University Press, 2007.

\bibitem[C74]{Chaitin}
Gregory J. Chaitin.
Information-theoretic limitations of formal systems.
\emph{Journal of the ACM}, 21(3):403--424, 1974.
\href{https://doi.org/10.1145/321832.321839}{doi:10.1145/321832.321839}.

\bibitem[FLL+26]{Fang}
Wen Fang, Xianxian Li, Jun Liu, Jie Luo, Yongxin Tong, and Ke Xu.
\emph{Self-Referential $K$-SAT and the Finite Analogue of G\"odel's Incompleteness Theorem}.
arXiv:2607.01671v3, 2026.
\url{https://arxiv.org/abs/2607.01671v3}.

\bibitem[G31]{Godel}
Kurt G\"odel.
\emph{\"Uber formal unentscheidbare S\"atze der Principia Mathematica und verwandter Systeme I}.
Monatshefte f\"ur Mathematik und Physik, 38:173--198, 1931.
\href{https://doi.org/10.1007/BF01700692}{doi:10.1007/BF01700692}.

\bibitem[GHY23]{GHY}
Balthasar Grabmayr, Volker Halbach, and Lingyuan Ye.
Varieties of self-reference in metamathematics.
\emph{Journal of Philosophical Logic}, 52:1005--1052, 2023.
\href{https://doi.org/10.1007/s10992-022-09696-y}{doi:10.1007/s10992-022-09696-y}.

\bibitem[HV14]{HV}
Volker Halbach and Albert Visser.
Self-reference in arithmetic I.
\emph{The Review of Symbolic Logic}, 7(4):671--691, 2014.
\href{https://doi.org/10.1017/S1755020314000288}{doi:10.1017/S1755020314000288}.

\bibitem[IP01]{ImpagliazzoPaturi}
Russell Impagliazzo and Ramamohan Paturi.
On the complexity of $k$-SAT.
\emph{Journal of Computer and System Sciences}, 62(2):367--375, 2001.
\href{https://doi.org/10.1006/jcss.2000.1727}{doi:10.1006/jcss.2000.1727}.

\bibitem[L69]{Lawvere}
F. William Lawvere.
Diagonal arguments and cartesian closed categories.
In \emph{Category Theory, Homology Theory and their Applications II},
Lecture Notes in Mathematics 92, pages 134--145. Springer, 1969.
Reprinted with author commentary in \emph{Reprints in Theory and Applications of Categories},
15:1--13, 2006.
\url{https://www.tac.mta.ca/tac/reprints/articles/15/tr15abs.html}.

\bibitem[L14]{Li}
Wei Li.
\emph{Mathematical Logic: Foundations for Information Science}.
Second Revised Edition. Progress in Computer Science and Applied Logic 25.
Springer Basel, 2014.
\href{https://doi.org/10.1007/978-3-0348-0862-0}{doi:10.1007/978-3-0348-0862-0}.

\bibitem[OA26]{OpenAIETH}
OpenAI.
\emph{The Exponential Time Hypothesis is False}.
Preprint, October 17, 2026. Revised October 25, 2026.
\href{https://github.com/openai/math/tree/main/preprints/The-Exponential-Time-Hypothesis-is-False-October-25-2026}{Revised preprint}.

\bibitem[PT19]{PT}
Andrei Popescu and Dmitriy Traytel.
A formally verified abstract account of G\"odel's incompleteness theorems.
In Pascal Fontaine, editor, \emph{Automated Deduction---CADE 27},
Lecture Notes in Computer Science 11716, pages 442--461. Springer, 2019.
\href{https://doi.org/10.1007/978-3-030-29436-6_26}{doi:10.1007/978-3-030-29436-6\_26}.

\bibitem[R36]{Rosser}
Barkley Rosser.
Extensions of some theorems of G\"odel and Church.
\emph{The Journal of Symbolic Logic}, 1(3):87--91, 1936.
\href{https://doi.org/10.2307/2269028}{doi:10.2307/2269028}.

\bibitem[VW18]{VassilevskaWilliams}
Virginia Vassilevska Williams.
On some fine-grained questions in algorithms and complexity.
In \emph{Proceedings of the International Congress of Mathematicians (ICM 2018)}, pages 3431--3472, 2018.
\href{https://doi.org/10.1142/9789813272880_0188}{doi:10.1142/9789813272880\_0188}.

\bibitem[Y03]{Yanofsky}
Noson S. Yanofsky.
A universal approach to self-referential paradoxes, incompleteness and fixed points.
\emph{The Bulletin of Symbolic Logic}, 9(3):362--386, 2003.
\url{https://arxiv.org/abs/math/0305282}.
\end{thebibliography}
\end{document}